\documentclass[11pt]{article}

\input xy
\xyoption{all}
\input epsf.tex
\usepackage{amssymb,amsmath}
\DeclareMathOperator{\im}{Im}

\DeclareMathOperator{\Ker}{Ker}
\DeclareMathOperator{\Vol}{Vol}
\DeclareMathOperator{\PD}{PD}
\DeclareMathOperator{\Gr}{Gr}
\usepackage{pstricks}
\usepackage{psfrag}
\usepackage{graphicx}
\usepackage[small,loose]{subfigure}
\usepackage[latin2]{inputenc}
\usepackage{epic}
\usepackage{latexsym}
\usepackage{mathrsfs}
\usepackage{bbm} % BlackBoeard letters
\usepackage{cancel} 
\usepackage{fleqn}

\def\harr#1#2{\smash{\mathop{\hbox to .3in{\rightarrowfill}}
 \limits^{\scriptstyle#1}_{\scriptstyle#2}}}

\def\ov{\overline}
\def\s2{\frac{1}{\sqrt2}}

\def\be{\begin{equation}}
\def\ee{\end{equation}}

\def\beqa{\begin{eqnarray}}
\def\eeqa{\end{eqnarray}}

\DeclareMathOperator{\Sq}{Sq}
\newcommand{\Hth}{\mathscr{H}_3}	% H_three

\def\Dsl{\,\raise.15ex\hbox{/}\mkern-13.5mu D} %can be subscripted
\def\d3{d^3}
\def\K{{\cal K}}
\def\W{{\cal W}}

\def\IR{\mathbb{R}}

\def\IZ{\mathbb{Z}}

\def\IS{\ensuremath{\mathbb{S}}}

\usepackage{a4wide}
\begin{document}

%\rightline{\tt hep-th/yymmnnn}
\rightline{CINVESTAV-MTY/07-025}
\rightline{RIKEN-TH-92}

\vfill

\begin{center}
{\Large \bf
Effects of brane-flux transition on black holes
in string theory}\\

\bigskip\bigskip\bigskip

{\large
        Oscar Loaiza-Brito$^a$ and 
        Kin-ya Oda$^b$}

\bigskip

{\it \small
        $^a$ Centro de Investigaci\'on y de Estudios Avanzados
                del I.P.N., Unidad Monterrey \\ 
        Cerro de las Mitras 2565, Col.\ Obispado, 64060, Monterrey, N.L., Mexico\\
        \smallskip
        {\rm E-mail: \tt oloaiza@fis.cinvestav.mx} 
%%%%%%%%%%%%%%%%%%%%%%%%%%%%%%%%%%%%%%%%%%%%%%%%%%%%%%%%%%%%%%%%%%%%%%%%%%%%
%   Why don't you abbreviate this address shorter? Answer: Not possible!!  %
%%%%%%%%%%%%%%%%%%%%%%%%%%%%%%%%%%%%%%%%%%%%%%%%%%%%%%%%%%%%%%%%%%%%%%%%%%%%

        \bigskip

        $^b$ Theoretical Physics Laboratory, RIKEN, Saitama 351-0198, Japan\\
        \smallskip
        {\rm E-mail: \tt odakin@riken.jp}
}

\bigskip\bigskip\bigskip

{\bf Abstract} \\
\end{center}

\begin{center} 
\begin{minipage}[h]{\textwidth} {
Extremal $\mathcal{N}=2$ black holes in four dimensions can be described by an ensemble of D3-branes wrapped on internal supersymmetric three-cycles of Calabi-Yau threefolds on which type IIB superstring theory is compactified.
We construct a similar configuration, with extra RR and NS-NS three-form fluxes being turned on. 
We can avoid the Freed-Witten anomaly on the D3-branes by enforcing  the pullback of these extra fluxes to the D3-branes to vanish at the classical level.
In the setup the D3-brane charge is not conserved since it is classified as a trivial class in twisted K-theory.
Consequently, the D3-branes may disappear by encountering an instantonic D5-brane localized in time.
We discuss what happens on the black hole described by such disappearing D3-branes, relying mainly on topological arguments.
Especially, we argue that another RR three-form flux will be left as a lump of remnant which is localized in the uncompactified four-dimensional space-time and that it may carry the same amount of degrees of freedom to describe a black hole, in cooperation with the original NS-NS flux, after this transition of the D3-branes. 
}
\end{minipage} 
\end{center}

%\leftline{CINVESTAV-FIS/02-027}
%\leftline{\tt hep-th/yymmnnn}

\vfill\vfill
% The above is simpler than below:
%\bigskip
%\bigskip
%\vspace{5cm}

%\leftline{February 2007}

\newpage

\section{Introduction}
Incorporation of extra three-form fluxes in string theory compactification has lead to exciting possibilities such as the stabilization of moduli and a possible explanation for the hierarchy problem, which provide new ways to construct more realistic phenomenological models~(for a review see~\cite{Grana:2005jc} and references therein). 
Besides the desirable consequences on the effective supergravity, the fluxes impose a stringent non-trivial topological constraint through the {\it Freed-Witten anomaly}~\cite{Freed:1999vc}.

This anomaly must be considered in any situation where we have a non-trivial NS-NS three-form flux. 
Essentially it forbids to wrap a D-brane on a cycle supporting a cohomologically non-trivial NS-NS flux in type II theories.
If one insists on wrapping a D-brane on this wrong cycle\footnote{Throughout this paper we shall consider spin$^c$ cycles only.}, the way to cancel the anomaly is to add magnetic sources for the gauge fields induced by the presence of the flux.
Those sources are provided by branes terminating on the anomalous brane, which is especially called {\it instantonic brane} when localized in time.
The appearance of instantonic branes carrying such an anomaly and its consequent cancellation by the addition of extra branes was firstly described in~\cite{Maldacena:2001xj}.

In the presence of non-trivial NS-NS flux~$H_3$, the instantonic brane triggers a transformation of D-branes into a coupling between NS-NS and RR fluxes.
What happens is that $N$~D$p$-branes may decay into the vacuum by encountering an instantonic D(p+2)-brane which supports $N$~units of NS-NS flux~$H_3$.
The D$p$-brane must be of codimension three in the Euclidean worldvolume of the instantonic brane. 
The charge of the disappeared D$p$-branes is, after the transition, carried by the coupling between the NS-NS flux~$H_3$ and the magnetic RR flux~$F_{6-p}$ related to the instantonic brane, such that the final flux configuration still carries the same quantum numbers as the disappeared D$p$-branes.

Roughly speaking, this topological process is a physical interpretation of the connection between integral cohomology and twisted K-theory~\cite{Witten:1998cd}.
In this connection, some non-trivial cohomology classes are obstructed to be lifted to twisted K-theory.
The obstruction follows precisely from the presence of the NS-NS flux~$H_3$. 
Hence, branes wrapped on cycles Poincar\'e-dual to the non-lifted forms are not BPS objects classifed by K-theory. 
The D$p$-branes which disappear by encountering the instantonic brane belong to $N$~torsion classes of K-theory.
Every configuration of branes wrapped on compact and non-trivial cycles in the presence of a NS-NS flux $H_3$ is potentially anomalous.
That implies that what seems to be a stable bunch of D$p$-branes can nevertheless be unstable to decay into vacuum by interacting with an instantonic brane.

D-branes wrapped on compact cycles have also been used in literature to describe extremal ${\cal N}=2$ black holes in four dimensions~\nocite{Strominger:1995cz, Suzuki:1995rt, Ooguri:2005vr, Ferrara:1995ih, Strominger:1996kf, Ferrara:1996dd, Ferrara:1996um, Behrndt:1996jn, Bertolini:1998se, Denef:2000nb, Ooguri:2004zv, Cardoso:2006cb, Pioline:2006ni}\cite{Strominger:1995cz}--\cite{Pioline:2006ni}.
They are constructed by wrapping D3-branes on three-cycles of a Calabi-Yau threefold into which type IIB theory has been compactified~\cite{Strominger:1995cz, Ooguri:2005vr}. All quantum numbers related to this BPS state, as well as the entropy, can be computed in terms of the internal symplectic geometry of the CY moduli space.

A natural question would be the stability of a black hole described by D3-branes when we turn on a suitable NS-NS flux~$H_3$ as in the above argument.\footnote{
	Recently, there also appeared several considerations on black holes under the presence of fluxes, for instance, the atractor mechanism in~\cite{Kallosh:2005bj, Kallosh:2005ax} and the possible influence of the presence of black holes on the moduli stabilization in~\cite{Danielsson:2006jg}.}
This is the issue we want to address in the present paper. 
We construct a configuration where one has a black hole described by D-branes, with an extra NS-NS flux $H_3$ being turned on.
We show that the D3-branes providing the description of a black hole may disappear and that they turn into a configuration of a remnant RR three-form flux~$F_3^\text{rem}$ which lives in the uncompactified four spacetime dimensions.
Our computation of the entropy is based on two important assumptions that the back-reaction from the fluxes in the initial configuration is negligible and that the flux-brane transition is reversible.

This paper is organized as follows: 
        In section~2 we review the construction of the extremal four-dimensional black hole by wrapping D3-branes on internal cycles. We present how to compute the black hole mass, entropy and effective potential in four dimensions.
        In section~3, we review the required conditions to preserve two supersymmetries in four dimensions. We construct a configuration where an extra internal flux is turned on without spoiling the supersymmetries. We examine the potential for the black hole in the flux background. 
        In section~4, we study the transition of the D3-branes into a remnant RR fluxes~$F_3^\text{rem}$ due to the appearance of an instantonic D5-brane.
        In section~5, this topological transformation process is interpreted with the black hole picture.
        In particular the black hole mass and entropy are estimated. 
		We also estimate the entropy of the fluxes~$F_3^\text{rem}$ into which the D3-branes has been transformed. 
        Finally some comments and open questions are addressed in the last section. 
In particular a black hole configuration with the remnant non-exact three-form flux $F_3^\text{rem}$ does not exist as a solution for the Einstein-Maxwell equations in four dimensions~\cite{SenGupta:2001cs}.
We show a possible interpretation that the ensemble of states described by the remnant flux~$F_3^\text{rem}$ could still be regarded as a black hole.
In Appendix A, we briefly review the connection between cohomology and K-theory by the so-called Atiyah-Hirzebruch Spectral Sequence. Appendix B is devoted to elucidate the brane-flux transition in terms of D-branes currents.

%---------------------------------------------------------------------------------------------------------------------------------------------------------------------------------------------------------------------------Section 2-----------------------------------------------
\section{Extremal black hole from wrapped D3-branes}

Consider a compactification of type IIB superstring theory on a Calabi-Yau three\-fold~$Y$. 
An extremal charged black hole of the effective ${\cal N}=2$ supergravity corresponds to D3-branes wrapped on internal special Lagrangian cycles of $Y$~\nocite{Strominger:1995cz, Suzuki:1995rt, Ooguri:2005vr, Ferrara:1995ih, Strominger:1996kf, Ferrara:1996dd, Ferrara:1996um, Behrndt:1996jn, Bertolini:1998se, Denef:2000nb, Ooguri:2004zv, Cardoso:2006cb, Pioline:2006ni}\cite{Strominger:1995cz}--\cite{Pioline:2006ni}.
By choosing a symplectic basis for the three-cycles in $Y$ as $\{A_I,B^I\}$ with $I$ running for $0,...,h^{2,1}(Y)$, let us write the cycle on which $N$ dyonic D3-branes are wrapped as
\begin{align}
\Sigma_3 = \sum_I (e^IA_I - m_IB^I),
\end{align}
where $e^I$ and $m_I$ are the number of times with which a single D3-brane is wrapped on the cycles $B^I$ and $A_I$, that is, the electric and magnetic charges for a single D3-brane, respectively. 
These cycles satisfy $A_I\cap A_J=B^I\cap B^J=0$ and $A_I\cap B^J=-B^J\cap A_I=\delta ^J_I$.
By using Poincar\'e-duality, a symplectic basis for 3-forms in $H^3(Y;\IZ)$ is defined as $\{\alpha_I,\beta^I\}$ where 
\begin{align}
\PD (A_I)&=\beta^I, &
\PD (B^I)&=\alpha_I,
\end{align}
such that 
\begin{align}
\int_{A_J}\alpha_I = -\int_{B^I}\beta^J=\delta^J_I.
\end{align}

The self-dual five-form $\mathscr{F}_5$, from which the electric D3-brane charge is computed, fulfills the following equation of motion~\cite{deAlwis:2006cb}
\begin{align}
d\ast \mathscr{F}_5=d\mathscr{F}_5=-\mu_3\kappa^2_0 \PD (\W_4)=\ast J_4
\label{Eq:G5}
\end{align}
where as usual, the self-duality has been imposed after getting the equations of motion.
The five-form $\mathscr{F}_5$ can be decomposed as~\cite{Suzuki:1995rt, Ooguri:2005vr, Denef:2000nb, Danielsson:2006jg}
\begin{align}
{\mathscr{F}_5}={\mathcal{F}}_3\wedge \omega_2 + \ast_6 {\mathcal{F}}_3\wedge\ast_4\omega_2
\end{align}
where $\omega_2$ is a two-form field strength in the non-compact four dimensions (through which the black hole's electric and magnetic charge are measured) and $\ast_4$ and $\ast_6$ are the Hodge dual in the indicated spaces. 
Note that $\alpha_I\wedge\beta^J=\delta_I^J\,\ast_61$.

Since the D3-branes wrapped on $\Sigma_3$ are BPS objects, it is expected that upon dimensional reduction from the ten-dimensional type IIB supergravity, they manifest theirselves as BPS objects in four-dimensions. Concretely, time-independent BPS configurations in four-dimensions are described by the metric~\cite{Denef:2000nb}
\begin{align}
ds^2= - e^{2U(\tau)}dt^2+\frac{ e^{-2U(\tau)}}{\tau^4}d\tau^2+\frac{ e^{-2U(\tau)}}{\tau^2}d\Omega^2,
\label{Eq:metric}
\end{align}
where $U(\tau)$ goes to zero for $\tau \rightarrow 0$ and to infinity at the horizon with $\tau = 1/r$.
By considering the above metric, $\mathscr{F}_5$ is self-dual provided
\begin{align}
\omega_2 & = g\sin\theta d\theta\wedge d\phi,	&
\ast_4\omega_2 &=  e^{2U}dt\wedge d\tau,
\end{align}
where $g$ is the elementary charge. Imposing the quantization conditions
\begin{align}
Nm^I&=\int_{A_I\times\IS^2}\mathscr{F}_5=2\pi g\int_{A_I}{\cal F}_3,	&
Ne_I&=\int_{B^I\times\IS^2}\mathscr{F}_5=2\pi g\int_{B^I}{\cal F}_3,
\end{align}
one finds that the internal component of the five-form $\mathscr{F}_5$ is written in terms of the symplectic 3-form basis as follows,
\begin{align}
{\cal F}_3 = N \sum_I (m^I\alpha_I - e_I\beta^I).
\end{align}
This form is supported on the internal three-cycle $\Pi_3$
\begin{align}
\Pi_3 = \sum_I (e^IA_I +m_IB^I).
\end{align}

According to the Dirac quantization condition, $\Sigma_3\cap \Pi_3=2m^I e_I\equiv 1$,
implying that the cycles are transversal to each other in $Y$. This fact leads us to consider some extra constrains in a model where D3-branes are wrapped on $\Sigma_3$ and extra flux has been turned on in the internal manifold, as we shall see in the following.
The first straightforward consequence of $\Sigma_3$ being transversal to $\Pi_3$ is that an extra NS-NS three-form supported on $\Sigma_3$ cannot be turned on, since it would carry the Freed-Witten anomaly~\cite{Freed:1999vc}. Since our goal is precisely to turn on extra flux, we will do it by selecting out a flux from the $h^{2,1}-1$ possible transversal cycles to $\Sigma_3$, which is precisely supported on $\Pi_3$. Such alignment between ${\cal F}_3$ and the extra flux will play an important role, as we shall see in Section 4.2.

\subsection{Black hole entropy}

Up to now we have reviewed how to express the self-dual five-form $\mathscr{F}_5$ in terms of internal symplectic forms. This notation also allows us to write down important black hole properties such as the mass and entropy. The BPS mass can be written in terms of a superpotential (resembling a ${\cal N}=1$ supergravity notation) as
\begin{align}
M^2_\text{BPS}&=e^{\K}|\W|^2,
\end{align}
where the superpotential $\W$ and the K\"ahler potential $\K$ are given by
\begin{align}
\W&=\int_{Y\times\IS^2}\mathscr{F}_5\wedge\Omega=\int_Y {\cal F}_3\wedge\Omega,&
\K&=- \log i\int_Y \Omega\wedge\ov\Omega.
\end{align}
and $\Omega$ being the holomorphic three-form in $Y$.
It is convenient to rewrite the above potentials in term of the phases  
\begin{align}
X^I=\int_{A_I}\Omega \qquad
\text{and} \qquad
F_I=\int_{B^I}\Omega,
\end{align}
from which $\Omega = X^I\alpha_I - F_I\beta^I$. Hence the superpotential reads 
\begin{align}
\W(X)&=N\left(e_IX^I-m^IF_I\right)
\end{align}
and the K\"ahler potential 
\begin{align}
e^{-\K(X,\ov{X})}&=2\left(\im \ov{\tau}_{IJ}\right) X^I\ov{X}^J,
\end{align}
where $\tau_{IJ}=\partial F_J/\partial X^I$.
Now the BPS mass is written as  
\begin{align}
M^2_\text{BPS}&=\frac{N^2}{2\left(\im \ov{\tau}_{IJ}\right)X^I\ov{X}^J}\left|e_IX^I+m^IF_I\right|^2.
\end{align}

%%% paragraph for entropy
Following the proposal by Ferrara and Kallosh~\cite{Ferrara:1996um}, the entropy of the black hole can be computed by extremizing the action
\begin{align}
S&=-\frac{\pi}{4}\left[ e^{-\K(X,\ov{X})} +2i\W(X) -2i\ov{\W}(\ov{X})\right]
\label{Eq:action}
\end{align}
with respect to $X_I$. 
The extremization condition for $S$ implies that (following the notation of~\cite{Ooguri:2005vr}) at the minimum,
\begin{align}
X_{e,m}^I&=\left(\frac{-iN}{\im \tau}\right)^{IJ}\left(e_J-m^K\ov{\tau}_{JK}\right),
\end{align}
for which the entropy is written as
\begin{align}
\mathcal{S}
        &= \frac{\pi}{2}\im\left[N^2
                \left(\frac{1}{\im \tau}\right)^{IK}
                \left(\frac{1}{\im \tau}\right)^{JL}
                \left(e_K-m^P\tau_{PK}\right)
                \left(e_L-m^Q\ov{\tau}_{QL}\right)\ov{\tau}_{IJ}\right]\nonumber\\
&\equiv N^2\mathcal{S}_\text{unit}. \label{entropy0}
\end{align}
Note that the entropy is proportional to $N^2$.

\subsection{Type IIB supergravity action}
The black hole potential $V_\text{BH}$ can be deduced from dimensional reduction from 10-dimensional type IIB supergravity action with the metric~(\ref{Eq:metric}).
The general bosonic part of the 10-dimensional action is given by
\begin{align}
S_\text{IIB}&=\int e^{-2\phi}\left(-\frac{1}{2}R\ast 1+2 d\phi\wedge\ast d\phi-\frac{1}{4}H_3\wedge\ast H_3\right)\nonumber\\
&\quad\mbox{}
-\frac{1}{2}\int\left( F_1\wedge\ast F_1+\widetilde{F}_3\wedge\ast \widetilde{F}_3+\frac{1}{2}\mathscr{F}_5\wedge\ast \mathscr{F}_5\right)-\frac{1}{2}\int C_4\wedge H_3\wedge F_3,
\end{align}
where
        $H_3=dB_2$, 
        $\widetilde{F}_3=F_3-C_0H_3$, and
        $\mathscr{F}_5=F_5-\frac{1}{2}C_2\wedge H_3+\frac{1}{2}B_2\wedge F_3$,
with
        $B_2$ being the NS-NS two-form potential, 
        $F_3=dC_2$ and $F_5=dC_4$ being the three and five-form RR field strengths, and 
        $C_0$ and $\phi$ being the RR and NS-NS scalars, respectively.

From this action, the equation of motion for the field strength~$\widetilde{F}_3$ is given by
\begin{align}
d\widetilde{F}_3=dF_3-F_1\wedge H_3,
\end{align}
where $F_1=dC_0$.
When compactifying on $Y$, the total ammount of D3-brane charge in $Y$  must be zero, namely $d\widetilde{F}_3|_Y=0$ and hence
\begin{align}
dF_3=F_1\wedge H_3.
\end{align}
This result strongly suggests that the current of D3-branes wrapped in internal 3-cycles of $Y$ are being absorbed by (or emanating from) a D5-brane localized in time that supports a non-trivial NS-NS flux.\footnote{A D5-brane cannot be wrapped in internal cycles since there are not five-cycles in $Y$.} 
We shall come back to this point later.

\subsection{Effective scalar potential for the Black Hole}

Let us now review how the self-dual five-form $\mathscr{F}_5$ (induced by the D3-branes) gives rise to a localized scalar potential in the uncompactified four-dimensional spacetime. We closely follow Refs.~\cite{Suzuki:1995rt, Denef:2000nb}.

In this section we consider the case where only the five-form flux $\mathscr{F}_5$ is turned on.
Then the above action reads
\begin{align}
S_\text{IIB}&= -\frac{1}{2\kappa_{10}^2}\int \mathscr{F}_5\wedge \ast \mathscr{F}_5\nonumber\\
& = -\frac{1}{4\kappa_{10}^2}\int \left( \omega_2\wedge \ast_4\omega_2\right)\int_{Y} \left({\cal F}_3 \wedge\ast_6{\cal F}_3\right).
\end{align}
By using the metric~(\ref{Eq:metric}) one finds that~\cite{Danielsson:2006jg}
\begin{align}
S_\text{IIB}= -\frac{1}{2\kappa_{10}^2}\int d(\text{Vol}_4)\frac{ e^{4U}}{r^4} V_\text{BH},
\end{align}
where
\begin{align}
V_\text{BH}= \int_Y {\cal F}_3\wedge \ast_6{\cal F}_3,
\end{align}
which can also been computed from
\begin{align}
V_\text{BH}=e^{\K}\left(D_i\W\,D_j\ov{\W}\,K^{ij}+|\W|^2\right).
\end{align}
Note that the effective scalar potential involving $V_\text{BH}$ is localized in the spacetime as expected.

\section{Turning on extra flux}
At this point we want to answer the question whether the presence of extra flux in the internal manifold would affect the black hole properties.
There are trivial cases in which the presence of three-form fluxes breaks some of the supersymmetries and we cannot have much control over black hole properties.\footnote{See~\cite{Saraikin:2007jc} for a treatment of non-supersymmetric black hole, which appeared after the first version of this manuscript.}
Here we aim at turning on fluxes while keeping $\mathcal{N}=2$ supersymmetry unbroken.

First of all, one cannot turn on NS-NS and RR fluxes such that there is a net D3-brane charge in the internal manifold. Although it is possible to cancel it by placing orientifold three-planes with negative tension and charge, its presence would spoil the ${\cal N}=2$ supersymmetry in the 4D background by projecting out the graviphotons in the vector multiplet which in turn give rise to the charge of the black hole (see for instance \cite{Grimm:2004uq}).\footnote{Note however that if one works at conifold singularities in the CY, ${\cal N}=2$ supersymmetry can indeed be maintained, as was shown in Refs.~\cite{Michelson:1996pn, Taylor:1999ii}.}
Second, as we have said, an extra  NS-NS flux cannot be supported on the same cycle where the D3-branes are wrapped, since the system would suffer from the Freed-Witten anomaly~\cite{Freed:1999vc}, for a review see~\cite{Evslin:2006cj} and references therein. Hence at least, the NS-NS flux to be turned on must be supported on $\Pi_3$. For the case we are interested in, a RR flux must be aligned to the NS-NS one, since otherwise would generate a tadpole.

These are the trivial cases we shall avoid. 
Below we consider a case in which three-form fluxes do not break ${\cal N}=2$ supersymmetry in the flat four-dimensional space time.

\subsection{Conditions to preserve ${\cal N}=2$ in 4D}

Turning on three-form fluxes and asking for the ${\cal N}=2$ supersymmetry to be preserved in four dimensions requires some more constraints, as was shown in~\cite{Curio:2000sc, Kachru:2004jr, Louis:2002ny}. Let us briefly review their main results.

Let us turn on $M$ units of the complex three-form flux $\Hth= F_3^\text{ini} +\tau H_3$, with $\tau$ as usual being the complex IIB dilaton: $\tau\equiv C_0+ie^{-2\phi}$. In terms of the symplectic basis, the initial RR and NS-NS three-form fluxes are written as
\begin{align}
F_3^\text{ini}
	&=	M_F \sum_I \left(p^I\alpha_I - q_I\beta^I\right),	&
H_3	&=	M_H \sum_I \left(m^I\alpha_I-e_I\beta^I\right).
\end{align}
Since we do not want to have a D3-brane charge contribution from the fluxes, we shall take that $H_3\wedge F_3^\text{ini}=0$, that is, $q_Im^I-p^Ie_I=0$. 
The complex three-form field strength reads
\begin{align}
\Hth = M \sum_I\left(P^I\alpha_I - Q_I\beta^I\right),
\end{align}
where
\begin{align}
P^I	&=	p^I+\tau m^I,	&
Q_I	&=	q^I+\tau e_I,
\end{align}
and $M=M_F+M_H$. The induced superpotential $\W_\text{flux}=\W_\text{NS}+\W_\text{R}$ can be written in terms of the symplectic forms as
\begin{align}
\W_\text{flux}= Q_I(\tau)X^I -P^I(\tau)F_I.
\end{align}
In the language of ${\cal N}=1$ supersymmetry, the gravitino mass matrix is~\cite{Curio:2000sc}
\begin{align}
\mathbf{S}_{AB} &= 
\left(\begin{array}{cc}
-\W_\text{flux}+2i\left(\im\tau\right)\W_\text{NS}&0\\
0&\W_\text{flux}
\end{array}\right).
\end{align}
Therefore, the conditions to preserve both of supersymmetries are to have $\W_\text{flux}=0$ and $\W_\text{NS}=0$, which is realized by treating the complex dilaton $\tau$ as a dynamical variable. Both gravitini remain massless if $\W_\text{flux}=\W_\text{NS}=0$ at the minimum. Hence, in order to preserve ${\cal N}=2$ supersymmetry, it is necessary to turn on both RR and NS-NS fluxes. Note also that we have taken the flux $H_3$ to be supported on $\Pi_3$ in order to avoid Freed-Witten anomalies.

Turning on the flux $\Hth$ induces an effective 4-dimensional scalar potential. From the ten-dimensional action the contribution from the flux is
\begin{align}
S_\text{4D}
	&=	-\frac{1}{4}\int_Y  e^{-2\phi}\left(H_3\wedge\ast H_3\right)
		-\frac{1}{2}\int_Y \Hth\wedge\ast \Hth
	 = 	-\int_Y V_\text{eff},
\end{align}
from which one obtains
\begin{align}
        V_\text{eff}
          = -\frac{1}{2}\sum_{IJKL}
                \left(C^2_0+\frac{e^{-2\phi}}{2}\right)
                \left(Q_I-{\cal M}_{IK}P^K\right)
                \left((\im {\cal M})^{-1}\right)^{IJ}
                \left(\ov{Q}_J-\ov{\cal M}_{JL}\ov{P}^L\right).
\end{align}
The negative definite matrix ${\cal M}$ is defined as in~\cite{Louis:2002ny}. 
Note that when the flux~$\Hth$ is turned on, the effective theory in four dimensions does not correspond to the usual ${\cal N}=2$ supergravity but to the {\it gauged} ${\cal N}=2$ supergravity, see for instance~\cite{Andrianopoli:1996cm}. 
The presence of the NS-NS flux~$H_3$ in the internal manifold induces a change in the effective action.
The resulting action of the gauged ${\cal N}=2$ supergravity in four dimensions is given in~\cite{Louis:2002ny}, where the ${\cal N}=2$ supersymmetry is maintained by the presence of a positive definite scalar potential. 

To summarize, both fluxes, RR and NS-NS, must be turned on and moreover, they must be supported on the same cycle $\Pi_3$.

\subsection{Black hole plus NS-NS flux}
Having described the effective localized scalar potetnial for the black hole, and the necessary conditions for the fluxes to be compatible with ${\cal N}=2$ supersymmetry, we now turn to the question about a system conforming to the above two issues: a black hole immersed in an scalar flux-potential. That is, D3-branes wrapped on internal cycles in the presence of internal three-form fluxes.

A configuration of D3-branes wrapped on~$\Sigma_3$ in the presence of the flux~$\Hth$ would effectively correspond to putting a 4D black hole in a background given by the scalar potential~$V_\text{eff}$.
The total effective 4D scalar potential (black hole plus flux) would be given as in~\cite{Danielsson:2006jg} by
\begin{align}
S\sim -\frac{1}{2\kappa^2_{10}}\int d(\Vol_4)\left(\frac{1}{r^4}V_\text{BH}+V_\text{eff}\right).
\end{align}
In the previous section we have seen that by wrapping D3-branes on internal three-cycles, we get an extremal ${\cal N}=2$ black hole in the four-dimensional effective theory when the extra flux~$\Hth$ is turned off. 
As is mentioned above, when one turns it on, the effective theory after compactification becomes a ${\cal N}=2$ gauged supergravity~\cite{Louis:2002ny}.

In order to compute black hole properties, we {\it assume that the BPS states described by the metric (\ref{Eq:metric}) are not altered by the presence of the scalar potential $V_\text{eff}$}. 
This assumption is backed up by the following argument.
Before turning on the extra flux $\Hth$, the five-form $\mathscr{F}_5$ related to the D3-brane is split into two parts: the internal $\mathcal{F}_3$ and the four dimensional divergent two-form $\omega_2$.
The former is compatible with BPS states in four dimensions.
The extra flux~$\Hth$ that we are turning on is precisely along the internal three form flux $\mathcal{F}_3$ from the D3-brane.
So we expect that such a BPS state does not suffer a change by turning on $\Hth$.

From our assumption, it follows that the black hole properties are not altered by the flux~$\Hth$ since the mass and charge associated to the black hole in four dimensions comes from the mass and charge for the internally wrapped D3-branes.
The flux~$\Hth$ does not contribute to the D3-brane charge or mass.
By construction, it merely changes the effective potential~$V_\text{eff}$ uniformly in four dimension. 
Specifically, since 
\begin{align}
\int_{A_I}\Hth &=Mm^I
\end{align}
is not a two-form in four dimensions, a four dimensional observer does not measure any divergence in the electromagnetic field (the scalar potential $V_\text{eff}$ is not localized as $V_\text{BH}$).
Also, the BPS mass is given in terms of the superpotential~$\W_\text{BH}$ which is null for the NS-NS flux.

On the other hand, we certainly expect to have a change in the entropy of the whole system, namely the black hole plus flux. 
Under our assumption, the total entropy of the system is given by
\begin{align}
  \mathcal{S}_\text{total}=\mathcal{S}_\text{BH}+\mathcal{S}_\text{flux},
  \end{align}
where $\mathcal{S}_\text{flux}$ is the entropy related to the presence of the NS-NS flux.
Note that reversibility is assumed here.
We note again that the presence of flux does not trigger a back-reaction since it does not contribute to the D3-brane charge tadpole.

\section{Flux-brane transition and twisted K-theory}

The configuration of branes and fluxes described in the previous sections leads to some non-trvial topological restrictions which in the end establishes a transition between branes and fluxes. In this section we shall see that such transition is properly understood in terms of twisted K-theory classes. Therefore we shall begin with a brief description of the role played by twisted K-theory trivial classes in the brane-flux transition. Later on, we shall show that our black hole configuration in the presence of extra flux cointains all the required ingredients to drive such a transition.

\subsection{Twisted K-theory and the MMS instanton}

In general, a homological cycle in a compact space can be wrapped by any D-brane, as far as the cycle is not trivial. However, this is not the only constriction to be fullfiled by the cycle. As it was shown in~\cite{Freed:1999vc}, the most general constriction reads,
\begin{align}
\W_3(\Sigma) + [H_3] =0,
\label{Eq:hFW}
\end{align}
where $\W_3(\Sigma)$ is the third Whitney class of a cycle $\Sigma$. 
If the cycle $\Sigma$ does not satisfy the above equation, a brane wrapped on it would suffer from the Freed-Witten anomaly. 
In the abscence of the NS-NS flux, the condition $\W_3(\Sigma)=0$ tells us that the cycle must be spin$^c$. 
Since we are working in a compact manifold, by the application of the Poincar\'e-duality, one can take the above homological constraint into a cohomological one. 
Let us be more specific and consider a $p$-cycle $\W_p$ in the compact space $M$ on which a D$p$-brane can be wrapped. 
We must take care on the required conditions for the cycle to be wrapped with a D-brane~\cite{Maldacena:2001xj}. 
The cycle defines a class $[\W_p]\in H_p(M;\IZ)$. 
By applying Poincar\'e-duality one relates the $p$-cycle $\Sigma_p$ with an $(n-p)$-cocycle $\sigma_{n-p}$ in $H^{(n-p)}(M;\IZ)$ where $\dim M =n$. 
Therefore, the cocycle $\sigma_{n-p}$ represents a D$p$-brane wrapped on $\Sigma_p$. Up to this point, D-brane RR charges are classified by (co)homology of $M$.

An extra condition, as we just saw, is that $\Sigma_p$ must fulfill  Eq.~(\ref{Eq:hFW}) which in cohomological form reads\footnote{We are considering the case for which there is not torsion.}
\begin{align}
d_3(\sigma_{n-p})
	&\equiv \Sq^3(\sigma_{n-p})+ [H_3]\wedge \omega_{n-p}=0,
\end{align}
where $\Sq^3$ is the Steenrod Square.

Hence one sees that the cocycle $\sigma_{n-p}$ represents an anomaly-free D$p$-brane provided it is closed under the linear map $d_3$. 
A D-brane wrapped on a non-trivial cycle but for which its representative cocycle is not closed under $d_3$ is not consistent and cannot be a stable object carrying a RR-charge. 
It is then expected that the group which classifies consistent D-brane cohomology representatives must contain cohomological non-trivial cocycles, which also belong to $\Ker d_3$. 

However, as well as for cohomology, it could be that some cocycles are not only exact under $d_3$ but also closed. 
The physical significance of these cocycles has been studied in the seminal paper by Maldacena, Moore and Seiberg (MMS)~\cite{Maldacena:2001xj}. 
A $d_3$-exact $(n-p)$-cocycle satisfy
\begin{align}
d_3(\pi_{n-p-3})= \Sq^3(\pi_{n-p-3})+[H_3]\wedge \pi_{n-p-3}= \PD(\Sigma_p\subset \Pi_{p+3})\wedge \pi_{n-p-3}\neq 0,
\end{align}
for $\pi_{n-p-3}$ being Poincar\'e-dual to the cycle $\Pi_{p+3}$.
The exactness under $d_3$ tells us that a D$p$-brane can be wrapped on $\Sigma_p$ since such a cycle satisfies the Freed-Witten anomaly-free condition. However they can be contained in a higher dimensional cycle $\Pi_{p+3}$ which in turn is not closed under $d_3$. Hence, D$p$-branes wrapped on $\Sigma_p$ can nevertheless be unstable since at some point will be immersed in a non-stable brane. Concretely, a D$p$-brane wrapping a spatial cycle $\Sigma_p$, propagates in time and terminates on an instantonic D$(p+2)$-brane wrapped on $\Pi_{p+3}$, for each unit of $H_3$.

Consequently, stable D-branes are represented by those cocycles which are closed but not exact under the map $d_3$. Let us define such a group by 
\begin{align}
E^p_3(M)= \frac{\Ker d_3|_{H^p}}{\im d_3|_{H^{p-3}}},
\end{align}
which roughly speaking is a ``cohomology'' of $d_3$. This algorithm by which integral cohomology is refined,  can actually be continued for higher linear maps such as $d_5$ that takes $p$-cocycles into $(p+5)$-cocycles and so on. 
The final set of cocycles, after a finite chain of steps, forms a group that resembles the so-called twisted K-theory group for $M$. 
A more formal description of this connection between cohomology and twisted K-theory, known as the Atiyah-Hirzebruch Spectral Sequence (AHSS), is given in appendix A. 
It is sufficient at this point to say that stable branes must be represented by twisted K-theory classes. 
An instantonic brane by itself is FW anomalous, but the anomaly is precisly cancelled by adding codimensional 3 D-branes wrapped on trivial homology classes under $d_3$.

It is important to point out a consequence of the decay of branes represented by trivial cycles in twisted K-theory. 
The appearence of the instantonic D$(p+2)$-brane induces the appearence of a flux $\ast G_{p+4}$ which couples magnetically to the instantonic brane. 
Therefore the charge of the dissapeared branes is now carried by the coupling between the NS-NS flux $H_3$ and the magnetic field strength $\ast G_{p+4}$. 
This is
\begin{align}
Q_{D_p}= \int H_3\wedge \ast G_{p+4}.
\end{align}

At the classical level and for branes wrapped on spin$^c$ cycles, as is the case for the present paper, the differential map $d_3$ is reduced to $d_3 = H_3\wedge $. 
A more pedestrian way to see the above issues is presented in Appendix B, where the flux-brane transition is described in terms of D-brane currents.

\subsection{D3-branes transition to flux}
Up to now we have considered wrapping D3-branes on $\Sigma_3$ in the presence of the flux $H_3=M(m^I\alpha_I-e_I\beta^I)$. Since the internal part of $\mathscr{F}_5$ is aligned with the flux $H_3$, one gets that

\begin{align}
{\cal F}_3 = H_3\wedge C_0,
\end{align}
where $C_0=N/M$. By constraining $C_0$ to be an integer, one can formally write that $C_0 \in H^0(Y;\IZ)$. In terms of the differential map $d_3$ the above expression reads
\begin{align}
d_3C_0=H_3\wedge C_0 \in H^3(Y;\IZ).
\end{align}
The zero-form $C_0$ is Poincar\'e-dual to a six-dimensional cycle in the Calabi-Yau manifold $Y$ and it is not closed under the linear map $d_3$. According to the previous statements, this means that a brane wrapped on the six-cycle (i.e., the whole CY) should be anomalous since the Freed-Witten condition is not fullfiled and $C_0$ would not belong to $E^0_3(Y)$. Therefore, an instantonic D5-brane wrapped on this cycle is in itself a potentially anomalous brane. 
On the other hand ${\cal F}_3$ is not only closed under $d_3$ but also exact and it is a trivial class in $E^3_3(Y)$
\begin{align}
{\cal F}_3 = d_3(C_0).
\end{align}

Hence, D3-branes wrapped on $\PD(-{\cal F}_3)=\Sigma_3$ would dissapear by encountering the instantonic D5-brane wrapped on $\Pi_6= \PD(C_0)$. However they are precisely the D3-branes giving rise to the black hole. Therefore, after the appearence of the instantonic D5-brane, the D3-branes disappear leaving as a remnant a magnetic flux $F_3^\text{rem}$ suppported on the uncompactified three-dimensional space.

Even if the decay of branes into flux is topologically possible, the transition occurs only when it is allowed energetically. 
A simple dimension counting argument shows that it is energetically favored to smear out a D3-brane into a D5 and turn it into a magnetic flux.\footnote{We thank J.~Evslin for bringing this point to our attention.}
However more qualitative description requires to use a specific metric as in~\cite{Polchinski:2000uf, Kachru:2002gs} where the deformed conifold metric is used to show that the nucleation of five-branes into which D3-branes are dissolved is energetically available. Since we are considering the topological aspects of the transition, we will not study the transition in a particular metric. The reader should keep in mind that even though the flux-brane transition is consistent with charge conservation in terms of the triviality of branes in twisted K-theory, such a decay will take place only if it is energetically allowed.

\section{Transition of black hole}
\label{trans_sec}
Let us consider $N$ D3-branes wrapped on $\Sigma_3$ and $M$ units of $\Hth$ in $\Pi_3$. 
A 4D observer measures the charge of the black hole (and an additional scalar field). 
Via the process described above, $M$ D3 branes disappear by encountering an instantonic D5-brane wrapped on the six-cycle in $Y$. 
After the disappearence of both D3 and D5-branes, there is a magnetic field strength remnant~$F_3^\text{rem}$ supported in the three-dimensional space.
Hence, the mass of the black hole is reduced since now there are only $(N-M)$ D3-branes wrapped on~$\Sigma_3$,
\begin{align}
M^2_\text{BPS}
        &= \frac{(N-M)^2}{2\left(\im \ov{\tau}_{IJ}\right)X^I\ov{X}^J}
           \left|e_IX^I+m^IF_I\right|^2.
\end{align}

\begin{figure}[btp]
\begin{center}
%\centering
%\epsfysize=8cm
\epsfysize=.3\textheight
%\leavevmode
%\epsfxsize= 9cm
\epsfbox{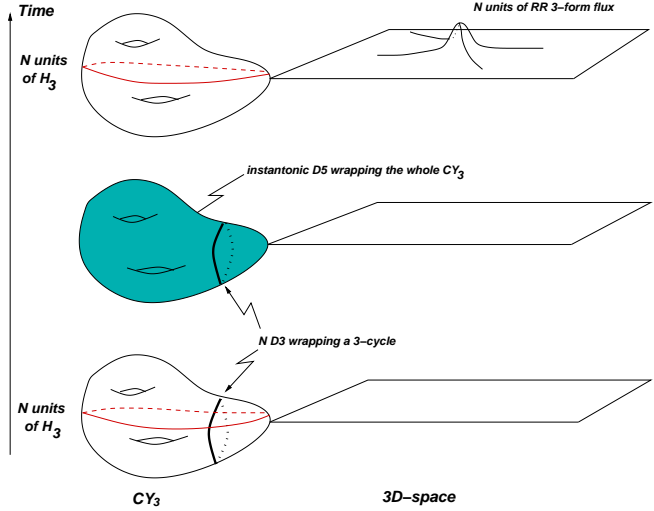}
\caption{Transformation of the black hole made from $N$ internally wrapped D3-branes into a configuration of a RR three-form flux in the uncompactified three-dimensional space. The transformation is triggered by the presence of the $N$~units of NS-NS flux and the appearence of an instantonic D5-brane.
        \label{Fig:BHdisapp}}
\end{center}
\end{figure}

The situation seems ``catastrophic''  for the black hole when there are the same number of D3-branes as NS-NS flux units ($N=M$). 
In such a case, the D3-branes are completely transformed into flux. 
The BPS mass vanishes $M_\text{BPS}^2=0$ and since the RR field strength $\mathscr{F}_5$ (for the D3-branes) has also disappeared, the charge cannot be measured by it.\footnote{
        The first case for a black hole to become massless was studied in~\cite{Strominger:1995cz} in the conifold approximation by shrinking to zero the cycle on which the D3-branes are supported.
        }
Hence, all the D3-brane charge is now carried by the NS-NS and RR fluxes, 
\begin{align}
Q_\text{D3}= \int_{\Pi_3\times\IR^3}H_3\wedge F_3^\text{rem}&=2\pi g N.
\end{align}
The whole process is depicted in Fig.~\ref{Fig:BHdisapp}.

The fact that the charge is now carried by the internal NS-NS flux~$H_3$ and the {\it new} RR flux~$F_3^\text{rem}$ created by the instantonic D5-brane lets us fix another property of~$F_3^\text{rem}$. 
Before the topological transition, the D3-brane charge was finite, equal to $N$ in D-brane charge units. 
After the transformation the brane charge is carried by the fluxes and henceforth must also be finite and equal to $N$. 
The $N$ units are given by the internal NS-NS flux, while 
\begin{align}
\int_{\IR^3}F_3^\text{rem}=2\pi g.
\end{align}
The RR flux $F_3^\text{rem}$ can be written as
\begin{align}
F_3^\text{rem}=F_{123}(\vec{x})\,dx^1\wedge dx^2\wedge dx^3.
\end{align}
Since the integral of $F_{123}(\vec{x})=F(\vec{x})$ over the three-dimensional space must be finite, $F(\vec{x})\rightarrow 0$ for $\vec{x}\rightarrow\infty$ and the three-flux must be localized in the three-dimensional space.
One simple solution would be something roughly of the form $F_3^\text{rem}=2\pi g \delta^3(r)\sin \theta dr\wedge d\theta\wedge d\phi$, being localized around the center of the former black hole. 
This is schematically expressed in the upper configuration in Fig.~\ref{Fig:BHdisapp}. 

A non-static solution is also permissible~\cite{SenGupta:2001cs}, e.g., a flux configuration which smears out over time. 
Such a situation would lead to more drastic consequence, \emph{decay of an extremal black hole}, which certainly requires more careful and detailed study. Here we leave it for future research.

Finally we must concern on the possibility that the final configuration of flux would not be consistent with ${\cal N}=2$ supersymmetry. Although a deeper study is required to fully answer this question, we can give here an explanation involving only the bosonic part of the fields. The initial configuration of flux, i.e., $\mathscr{F}_5= \omega_2\wedge {\cal F}_3$ carry the electric and magnetic charges of the black hole, being the two-form $\omega_2$ divergent. In other words, the equation of motion for the five-form field strength is
\begin{align}
d\mathscr{F}_5= d\omega_2\wedge {\cal F}_3.
\end{align}
This is actually an exact 6-form in the ten-dimensional spacetime. After the transition, such a form disappears leaving a remnant which indeed can built up a six-form from which the same previous-transition black hole charge is gathered. Since the produced RR field strength $F_3^\text{rem}$ satisfies the Bianchi identity $dF_3^\text{rem}=0$, can be locally written as an exact form, this is $F_3^\text{rem}=dC_2$. Hence the final configuration of fluxes consists of an internal three-form $H_3$ (playing the role of the former ${\cal F}_3$) and a divergent two-form $C_2$ (plying the role of the former $\omega_2$), where $C_2= g\sin\theta d\theta\wedge d\phi$. In this way, $C_2$ is locally a solution for the black hole equations for the metric (\ref{Eq:metric}).

\subsection{Entropy after transition}
So far, we have shown how the mass and charge are transformed by the appearance of the instantonic brane. However, as we said before, the entropy also suffer from changes by the presence of external fluxes.
We aim at computing the entropy of the configuration after the transformation between branes and fluxes.
By assuming that the process is reversible, it is possible to compute the entropy related to the flux $H_3\wedge F_3^\text{rem}$, which lives both in the uncompactified three-dimensional space and in the internal space $Y$. First of all, it is clear that for the remaining $(N-M)$ D3-branes, the black hole entropy reads,
\begin{align}
\mathcal{S}_\text{BH}^\text{after}&=(N-M)^2 \mathcal{S}_\text{unit}.
\end{align}
Now the $M$ units of D3-brane charge, carried by the fluxes $H_3\wedge F_3^\text{rem}$, should contribute with an entropy $\mathcal{S}_{\text{flux}}$ such that
\begin{align}
\mathcal{S}'= \mathcal{S}_\text{RR}+\mathcal{S}_\text{flux},
\end{align}
since both fluxes are localized in diferent spaces.
We have assumed that the process is reversible and hence the total entropy is conserved. 
Then it follows that the entropy of the black hole before the transition should be equal to the entropy of the lowered-mass black hole plus the RR flux
\begin{align}
\mathcal{S}^{\text{before}}_{\text{BH}}&=\mathcal{S}^{\text{after}}_{\text{BH}}+\mathcal{S}_{\text{RR}}.
\end{align}
The entropy for the remnant RR flux is hence given by
\begin{align}
\mathcal{S}_{\text{RR}}&=M(2N -M)\mathcal{S}_\text{unit}.
\end{align}
Note that for the case~$N=M$, all entropy is related now to the RR flux.

One may think that any flux~$F_3^\text{rem}$ could give rise to a static spherical solution for Einstein-Maxwell (EM) equations that again reproduces a black hole in the four-dimensional effective theory.
If this was the case, the difference between the original black hole and the one produced by the flux would only be the type of associated degrees of freedom. 
However, as was shown in~\cite{SenGupta:2001cs} a direct solution for a globally non-exact form is not possible, namely, there cannot be any spherically symmetric charged black hole solution for the EM equations under the presence of the three-form flux $F_3^\text{rem}$.
In this sense, a three-form flux cannot give rise to a black hole unless it is globally exact.\footnote{In~\cite{Rahaman:2006qm} a wormhole solution is studied with the same ingredients.}

If our assumption that the process being reversible should hold, the final flux configuration must somehow describe the same black hole.
The following points can be made to support this argument.
\begin{itemize}
        \item Even when the BPS-condition does not hold with non-zero $\Hth$ flux, the initial D3-branes would still represent a black hole due to the Thorne's Hoop conjecture. Also the same argument seems to conclude that the final bunch of RR flux in 4D becomes a black hole.
        \item Regardless of the assumptions, it would be natural that the equivalence of brane and flux holds in general not only for the RR sector but also for the NS-NS sector including gravity.
        \item The created RR flux is also localized in the three-dimensional space. 
        \item If the three-form remnant flux would have been a NS-NS flux, which directly involves gravity, the conclusion that such a flux describes a black hole would be more naturally expected. 
A transition between D3-branes and NS-NS fluxes can occur for instantonic NS5-branes in the presence of non-trivial units of RR flux. 
This process is actually an S-dual picture of the flux-brane transition of our consideration, which has been sucesfully tested for particular cases (see~\cite{Evslin:2006cj} and references therein).
        \item It might be that the resultant ensemble of states consisting of the flux is still a black hole even though there cannot be a direct black hole solution with flux~\cite{Mathur:2005zp} (see below).
\end{itemize}

\section{Conclusion and Discussions}
\label{conc_sec}
In this paper we have shown that, under the presence of the non trivial NS-NS three-form flux~$H_3$ in the internal space, the four dimensional black hole described by D3-branes may disappear via the topological process that transforms branes into fluxes.
A NS-NS flux induces the Freed-Witten anomaly on an instantonic D5-brane which must be canceled by a D3-brane ending on it.
Applying this fact to the system of the D3-branes wrapped on an internal Calabi-Yau cycle under the presence of extra NS-NS flux~$H_3$, we have shown that the black hole described in four dimensions would suffer from the same topological transition.

        Under the four-dimensional perspective, the black hole would disappear leaving as a remnant a RR three-form flux~$F_3^\text{rem}$ localized in the uncompactified four spacetime dimensions. 
        In order to compute the black hole quantities, we have assumed that the black hole represents a BPS state even in the presence of the NS-NS flux~$H_3$, namely, that there is not interplay between the effective scalar potential induced by the presence of the fluxes and the extremal black hole except for the transition drived by the instantonic brane. 
        In other words, our assumption is that the BPS states in the usual ${\cal N}=2$ four-dimensional supergravity are preserved in the ${\cal N}=2$ gauged supergravity.
The charge (and mass) carried by the black hole before the topological transition driven by the instantonic D5-brane is afterwards carried by the coupling between the NS-NS flux $H_3$ and another RR three-form flux $F_3^\text{rem}$ emanating from the D5-brane.

If our assumption of the BPS-ness of the initial black hole is valid, it implies that even the extremal black hole suffers from the transition that exchanges topologically different configurations which carry the same charge and mass.

Regardless of the extremality of the initial black hole, the solution to the Einstein-Maxwell equations in the presence of the remnant RR flux~$F_3^\text{rem}$, which is not globally exact, cannot be a spherically symmetric charged black hole solution, implying that at the level of gravity the initial black hole does suffer a change.
However, this does not necessarily mean that the final flux configurations cannot make an (extremal) black hole.
Recently, it has been conjectured that an ensemble of the different solutions without horizon corresponds to a black hole~\cite{Mathur:2005zp}. 
According to it, the horizon radius is nothing but the size of the region where the solutions in the ensemble differs each other.
This might be the case for the final configurations. Interesting enough, the wormhole solution is found in~\cite{Rahaman:2006qm} with which the integration of~$F_3^\text{rem}$ over a three-cycle can be finite and equal to the charge of the disappeared black hole.

There are many open questions.
First of all, the already mentioned topological transformation was explored some years ago in the context of the conifold approximation~\cite{Greene:1995hu}.
In this scheme, the moduli space describing the geometry of the conifold on which a D3-brane is wrapped changes after a topological transformation under which the three-cycle of the confiold shrinks to zero and blow up into a two-cycle. Particularly, the complex structure moduli were exchanged into the K\"ahler moduli. It would be very interesting to go further and study what happens with the Calabi-Yau moduli under the presence of an instantonic brane.

In this paper, we have treated the purely topological aspects of the process and therefore no dynamics is analyzed.
Time evolution of the S(pacelike)-brane has already been considered for simpler setup~\cite{Hashimoto:2002sk,Hashimoto:2003qx}.
It is of importance to study the dynamics of this transition from black hole into the RR flux.

Another interesting question arises from the computation of the entropy. As was mentioned at the begining of this paper, the black hole entropy can be computed by extremizing the central charge (which turns out to be also an extremization of the superpotential).
This is now a standard procedure to compute it, since we are dealing basically with open strings (with D3-branes present). 
Now, since we are addressing only a topological transition, we expect to have a configuration keeping the same quantum numbers and more importantly the same entropy. 
However, due to the fact that for $N=M$ we do not have D3-branes anymore, there are not degrees of freedom related to open strings. The entropy should be computed from a closed string perspective. This strongly suggests a way to compute entropy by other methods as was conjectured in~\cite{Ooguri:2004zv}, where the entropy is gathered from the topological partition function $|Z_\text{top}|^2$. It would be very interesting to have a detailed description of the entropy for a NS-NS field in terms of the topological partition function, which would provide another support for this conjecture.

\bigskip

\begin{center}
{\bf Acknowledgments}
\end{center}
\noindent We are grateful to Tsuguhiko Asakawa for the careful reading of the manuscript and to Amihay Hanany and Kazutoshi Ohta for helpful suggestions and discussions.
We also appreciate
        Daniel Cremades,
        Jarah Evslin,
        Hugo Garcia-Compean, 
        Koji Hashimoto, 
        Andrei Micu, 
        Gianmassimo Tasinato,
        Seiji Terashima, and
        the referee
        for useful comments.
We thank Manuel Drees and Hans-Peter Nilles for the support at Bonn University where this project began. O.~L.-B.\ also thanks RIKEN Theoretical Physics Laboratory for the hospitality.
O.~L.-B.\ is partialy supported by CONACyT (Mexico) under the program ``repatriaci\'on.''
The work of K.O.\ is partialy supported by the Special Postdoctoral Researchers Program at RIKEN. 

\bigskip
%\newpage 

\appendix
\begin{flushleft}
{\bf \Large Appendix}
\end{flushleft}
\section{The Atiyah-Hirzebruch Spectral Sequence}

Here we briefly review the connection between cohomology and K-theory in terms of the Atiyah-Hirzebruch Spectral Sequence (AHSS) (for further details, see~\cite{Diaconescu:2000wy, Maldacena:2001xj, Evslin:2006cj}). Essentially AHSS is an algorithm which relates integral cohomology with twisted K-theory.

The relation involves the construction of twisted K-theory classes from integral cohomology classes by a finite number of steps.  
In general, an integral cohomology class $[\omega_p]\in H^p(X;\IZ)$ does not come from a twisted K-theory class $[x]\in K(X)$. Hence, the  algorithm begins with cohomology. At this step, cohomology is the first approximation to twisted K-theory and it is denoted by $E_1(X)$.
At the $m$-th step, the approximate group is denoted by $E^p_m(X)$, where\footnote{In this section, ``$\im$'' stands for an image rather than an imaginary part.}
\begin{align}
K(X)\sim E^p_m \equiv \frac{\Ker d_m|_{E^p_{m-2}}}{\im d_m|_{E^{p-m}_{m-2}}},
\end{align}
such that, the first approximation is
\begin{align}
E^p_1(X) = \oplus_p H^p(X;\IZ).
\end{align}

The second step is to consider forms which are closed under the differential map
\begin{align}
d^3 \equiv \Sq^3 + H_3\wedge,
\end{align}
with $d^3$: $H^p(X;\IZ) \rightarrow H^{p+3}(X;\IZ)$. Discarding those which are exact, this defines the group
\begin{align}
E^p_3(X)=\frac{\Ker d_3|_{H^p}}{\im d_3|_{H^{p-3}}}.
\end{align}
After this step, only those forms which are closed will survive and represent stable D-branes in string theory provided the NS-NS field is identified with $H_3$. 
Those forms which are not closed represent the instantonic branes that we discuss in this paper. 
Finally, $(p+3)$-forms which belong to the trivial class satisfy
\begin{align}
d^3 \omega_p = \Sq^3(\omega_p) +  H_3\wedge\omega_p=\sigma_{p+3},
\end{align}
and represent branes which can be unstable. 
Note that such forms belong to torsion classes in K-theory according to the integral class of $H_3$ (upon the isomorphism with the field).

One can go further defining several groups in order to get a closer approximation to K-theory. However, the algorithm ends after a finite number of steps. At the end, one gets a group which is called ``associated graded group'' $\Gr(X)$ given by,
\begin{align}
\Gr(K_H(X))=\oplus_p E^p(X).
\end{align}
This group is in some cases the K-theory group.\footnote{In the context of string theory, this happens in the absence of orientifold planes.} 
However, in other cases it is necessary to solve an extension problem, since
\begin{align}
\Gr(K_H(X))=\oplus_p K_{H,p}(X)/ K_{H, p+1}(X).
\end{align}
In cases as in this paper, the second step is the final approximation. Hence, by dropping all forms closed under $d^3$ one gets a K-theory class. All branes which belong to a non-trivial class in this ``cohomology'' of $d^3$ are stable objects in string theory.

%A particular case in type IIB string theory compactification on a Calabi-Yau $Y$ arises by applying the AHSS to zero and three-forms. 

\section{Flux-brane transition}
The configuration of $N$ D3-branes wrapped on $\Sigma_3$ and an extra NS-NS flux on $\Pi_3$ suffers from a topological transition~\cite{Maldacena:2001xj, Evslin:2006cj, Loaiza-Brito:2006se}. 
To see this in more physical terms, consider the D3-brane action in 10~dimensions,\footnote{We shall use $F_3$ to denote the RR flux which is the magnetic field strength for a D5-brane and $G_3$ as the internal part of the self-dual five-form $G_5$.}
\begin{align}
S&=-\frac{1}{2\kappa^2_0}\int \mathscr{F}_5\wedge\ast \mathscr{F}_5+\mu_3\int C_4\wedge \PD (\W_4)+\frac{1}{4\kappa^2_0}\int C_4\wedge H_3\wedge F_3,
\end{align}
where the last term is the Chern-Simons term  from the ten-dimensional type IIB supergravity. 
Now the equation of motion for the self-dual $\mathscr{F}_5$ is modified from Eq.~\eqref{Eq:G5} as~\cite{deAlwis:2006cb}
\begin{align}
d\ast \mathscr{F}_5&=-\kappa^2_0\mu_3\,\PD (\W_4)+H_3\wedge F_3.
\end{align}
Note that the 3-form fluxes~$H_3$ and~$F_3$ are supported transversally to the cycle~$\Sigma_3$.

The coupling~$H_3\wedge F_3$ contributes to the D3-brane charge and this is usually used in flux compactifications. 
However if $dF_3\neq 0$, the presence of a D5-brane is required and there are only two possible ten-dimensional configurations:
\begin{enumerate}
\item
There is a normal D5-brane on which the D3-brane is attached. 
The intersection submanifold is of codimension three on~$\W_6$. 
Nevertheless, since there are not five-cycles in~$Y$ for a D5-brane to be wrapped, this configuration must be discarded for our model.
\item
There is an instantonic D5-brane localized in time (D5 wrapped on a six-dimensional cycle on~$Y$) upon which the D3-brane terminates. A D3-brane disappears for each unit of NS-NS flux~$H_3$ by the topological transition.
\end{enumerate}
Let us study the second option.
The equations of motion implies the non-conservation of the D3-brane current
\begin{align}
d\ast J_4=H_3\wedge dF_3,
\end{align}
where the D3-brane charge is computed through~$\ast J_4$.
The coupling~$H_3\wedge F_3$ induces a D3-brane charge but it does not represent a presence of D3-branes.
What it is conserved is the total charge from D3-branes and fluxes
\begin{align} 
d(\ast J_4+F_3\wedge H_3)=0.
\end{align}
First in the absence of D5-branes, or equivalently for $F_3=0$, the only source for D3-brane charge is the $N$ D3-branes themselves which is measured by $\ast J_4$. 
At a certain time, an instantonic D5-brane appears supporting $M$ units of NS-NS flux. 
In principle, the brane is anomalous in the sense of Freed-Witten~\cite{Freed:1999vc}. However the incoming $N$ D3-branes cancel the anomaly~\cite{Maldacena:2001xj}, by ending at the instantonic brane.

Since the variation of the current of D3-branes is
\begin{align}
\int d\ast J_4&=M,
\end{align}
there remains only $N-M$ D3-branes after the instantonic disappearance.
The $M$ D3-branes have disappeared by contacting the instantonic brane.
However, there is a remnant which corresponds to a magnetic field strength related to the instantonic D5-brane. This flux is a 3-form $F_3$ {\it supported on a transversal cycle to the instantonic brane}. The D3-brane charge is now carried by the coupling 
\begin{align}
          \int_{\Pi_3\times\IR^3}H_3\wedge F_3
        = M\cdot 1 
        = M,
\end{align}
satisfying the fact that the total charge is conserved.

The equation of motion for the flux~$\mathscr{F}_5$ can be reduced from the ten-dimensional spacetime to the compact CY manifold~$Y$ as
\begin{align}
d\ov\ast \mathscr{F}_5&=\ov\ast J_4+H_3\wedge\ov\ast G_7,
\end{align}
where $\ov\ast$ is the Hodge dual acting on the compact manifold~$Y$ and time so that $\ov\ast G_7$ is a zero-form in~$Y$ and a three-form in the un\-compactified three-dimensional space.
The flux~$\ov\ast G_7=\widetilde{F}_0$ corresponds to the magnetic field strength for {\it one} instantonic D5-brane wrapped on the six-cycle in~$Y$.\footnote{
        Such a brane is effectively seen as an instanton in 4D. Although we shall say nothing about it, the appearence of the instantonic brane would be seen by an observer in 4D as the interaction of a black hole with an instanton.} 
Hence, before it appears, the flux satisfy $\ov\ast G_7=0$ and the D3-brane charge is measured by 
\begin{align}
Q^\text{before}&=\int\ov\ast J_4= \int_{\Pi_3\times\IS^2}\mathscr{F}_5=N.
\end{align}
The variation of the current implies that after the appearance of the instantonic brane,
\begin{align}
Q^\text{after}&=\int\ov\ast J_4=\int_{\Pi_3\times\IS^2}\mathscr{F}_5=N-M,
\end{align}
from which we conclude that $M$~D3-branes have disappeared and been transformed into the flux~$H_3\wedge \widetilde{F}_0$. 
%This topological transformation have an interesting effect for a four-dimensional observer, as we shall see.

%Notice that the lost charge is now carried by 
%\begin{align}
%\int_{\Pi_3\times\IR^3}H_3\wedge F_3=\int_{\Pi_3} H_3\wedge\widetilde{F}_0=M\cdot 1=M.
%\end{align}
%where $F_3$ is the RR field strength for the instantonic D5-brane. This three-form flux must be supported transversally to the position of the instantonic brane, which in the present configuration means that it is located in the three-dimensional space.

%\begin{itemize}
        %\item Did we clearly state that the initial RR flux disappears after the transition?
        %\item Does the instantonic D5-brane appear as a point in the uncompactified 4-d space or as a uniform constant in 4-d? Is it stated clearly?
  %\end{itemize}

\bibliography{bhd}

\providecommand{\bysame}{\leavevmode\hbox to3em{\hrulefill}\thinspace}
\begin{thebibliography}{10}

\bibitem{Grana:2005jc}
M.~Grana, \emph{{Flux compactifications in string theory: A comprehensive
  review}}, Phys. Rept. \textbf{423} (2006), 91--158,  \texttt{hep-th/0509003}.

\bibitem{Freed:1999vc}
D.~S. Freed and E.~Witten, \emph{{Anomalies in string theory with D-branes}},
  (1999),  \texttt{hep-th/9907189}.

\bibitem{Maldacena:2001xj}
J.~M. Maldacena, G.~W. Moore, and N.~Seiberg, \emph{{D-brane instantons and
  K-theory charges}}, JHEP \textbf{11} (2001), 062,  \texttt{hep-th/0108100}.

\bibitem{Witten:1998cd}
E.~Witten, \emph{{D-branes and K-theory}}, JHEP \textbf{12} (1998), 019,
  \texttt{hep-th/9810188}.

\bibitem{Strominger:1995cz}
A.~Strominger, \emph{Massless black holes and conifolds in string theory},
  Nucl. Phys. \textbf{B451} (1995), 96--108,  \texttt{hep-th/9504090}.

\bibitem{Suzuki:1995rt}
H.~Suzuki, \emph{{Calabi-Yau compactification of type IIB string and a mass
  formula of the extreme black holes}}, Mod. Phys. Lett. \textbf{A11} (1996),
  623--630,  \texttt{hep-th/9508001}.

\bibitem{Ooguri:2005vr}
H.~Ooguri, C.~Vafa, and E.~P. Verlinde, \emph{{Hartle-Hawking wave-function for
  flux compactifications}}, Lett. Math. Phys. \textbf{74} (2005), 311--342,
  \texttt{hep-th/0502211}.

\bibitem{Ferrara:1995ih}
S.~Ferrara, R.~Kallosh, and A.~Strominger, \emph{N=2 extremal black holes},
  Phys. Rev. \textbf{D52} (1995), 5412--5416,  \texttt{hep-th/9508072}.

\bibitem{Strominger:1996kf}
A.~Strominger, \emph{{Macroscopic Entropy of $N=2$ Extremal Black Holes}},
  Phys. Lett. \textbf{B383} (1996), 39--43,  \texttt{hep-th/9602111}.

\bibitem{Ferrara:1996dd}
S.~Ferrara and R.~Kallosh, \emph{{Supersymmetry and Attractors}}, Phys. Rev.
  \textbf{D54} (1996), 1514--1524,  \texttt{hep-th/9602136}.

\bibitem{Ferrara:1996um}
S.~Ferrara and R.~Kallosh, \emph{{Universality of Supersymmetric Attractors}},
  Phys. Rev. \textbf{D54} (1996), 1525--1534,  \texttt{hep-th/9603090}.

\bibitem{Behrndt:1996jn}
K.~Behrndt et~al., \emph{{Classical and quantum N = 2 supersymmetric black
  holes}}, Nucl. Phys. \textbf{B488} (1997), 236--260,
  \texttt{hep-th/9610105}.

\bibitem{Bertolini:1998se}
M.~Bertolini, P.~Fre, R.~Iengo, and C.~A. Scrucca, \emph{{Black holes as
  D3-branes on Calabi-Yau threefolds}}, Phys. Lett. \textbf{B431} (1998),
  22--30,  \texttt{hep-th/9803096}.

\bibitem{Denef:2000nb}
F.~Denef, \emph{Supergravity flows and d-brane stability}, JHEP \textbf{08}
  (2000), 050,  \texttt{hep-th/0005049}.

\bibitem{Ooguri:2004zv}
H.~Ooguri, A.~Strominger, and C.~Vafa, \emph{Black hole attractors and the
  topological string}, Phys. Rev. \textbf{D70} (2004), 106007,
  \texttt{hep-th/0405146}.

\bibitem{Cardoso:2006cb}
G.~L. Cardoso, V.~Grass, D.~Lust, and J.~Perz, \emph{{Extremal non-BPS black
  holes and entropy extremization}}, JHEP \textbf{09} (2006), 078,
  \texttt{hep-th/0607202}.

\bibitem{Pioline:2006ni}
B.~Pioline, \emph{Lectures on on black holes, topological strings and quantum
  attractors}, Class. Quant. Grav. \textbf{23} (2006), S981,
  \texttt{hep-th/0607227}.

\bibitem{Kallosh:2005bj}
R.~Kallosh, \emph{Flux vacua as supersymmetric attractors},  (2005),
  \texttt{hep-th/0509112}.

\bibitem{Kallosh:2005ax}
R.~Kallosh, \emph{New attractors}, JHEP \textbf{12} (2005), 022,
  \texttt{hep-th/0510024}.

\bibitem{Danielsson:2006jg}
U.~H. Danielsson, N.~Johansson, and M.~Larfors, \emph{Stability of flux vacua
  in the presence of charged black holes}, JHEP \textbf{09} (2006), 069,
  \texttt{hep-th/0605106}.

\bibitem{SenGupta:2001cs}
S.~SenGupta and S.~Sur, \emph{Spherically symmetric solutions of gravitational
  field equations in kalb-ramond background}, Phys. Lett. \textbf{B521} (2001),
  350--356,  \texttt{gr-qc/0102095}.

\bibitem{deAlwis:2006cb}
S.~P. de~Alwis, \emph{Transitions between flux vacua},  (2006),
  \texttt{hep-th/0605184}.

\bibitem{Saraikin:2007jc}
K.~Saraikin and C.~Vafa, \emph{Non-supersymmetric black holes and topological
  strings},  (2007),  \texttt{hep-th/0703214}.

\bibitem{Grimm:2004uq}
T.~W. Grimm and J.~Louis, \emph{The effective action of n = 1 calabi-yau
  orientifolds}, Nucl. Phys. \textbf{B699} (2004), 387--426,
  \texttt{hep-th/0403067}.

\bibitem{Michelson:1996pn}
J.~Michelson, \emph{{Compactifications of type IIB strings to four dimensions
  with non-trivial classical potential}}, Nucl. Phys. \textbf{B495} (1997),
  127--148,  \texttt{hep-th/9610151}.

\bibitem{Taylor:1999ii}
T.~R. Taylor and C.~Vafa, \emph{{RR flux on Calabi-Yau and partial
  supersymmetry breaking}}, Phys. Lett. \textbf{B474} (2000), 130--137,
  \texttt{hep-th/9912152}.

\bibitem{Evslin:2006cj}
J.~Evslin, \emph{{What does(n't) K-theory classify?}},  (2006),
  \texttt{hep-th/0610328}.

\bibitem{Curio:2000sc}
G.~Curio, A.~Klemm, D.~Lust, and S.~Theisen, \emph{On the vacuum structure of
  type ii string compactifications on calabi-yau spaces with h-fluxes}, Nucl.
  Phys. \textbf{B609} (2001), 3--45,  \texttt{hep-th/0012213}.

\bibitem{Kachru:2004jr}
S.~Kachru and A.-K. Kashani-Poor, \emph{Moduli potentials in type iia
  compactifications with rr and ns flux}, JHEP \textbf{03} (2005), 066,
  \texttt{hep-th/0411279}.

\bibitem{Louis:2002ny}
J.~Louis and A.~Micu, \emph{{Type II theories compactified on Calabi-Yau
  threefolds in the presence of background fluxes}}, Nucl. Phys. \textbf{B635}
  (2002), 395--431,  \texttt{hep-th/0202168}.

\bibitem{Andrianopoli:1996cm}
L.~Andrianopoli et~al., \emph{{N = 2 supergravity and N = 2 super Yang-Mills
  theory on general scalar manifolds: Symplectic covariance, gaugings and the
  momentum map}}, J. Geom. Phys. \textbf{23} (1997), 111--189,
  \texttt{hep-th/9605032}.

\bibitem{Polchinski:2000uf}
J.~Polchinski and M.~J. Strassler, \emph{The string dual of a confining
  four-dimensional gauge theory},  (0300),  \texttt{hep-th/0003136}.

\bibitem{Kachru:2002gs}
S.~Kachru, J.~Pearson, and H.~L. Verlinde, \emph{Brane/flux annihilation and
  the string dual of a non- supersymmetric field theory}, JHEP \textbf{06}
  (2002), 021,  \texttt{hep-th/0112197}.

\bibitem{Rahaman:2006qm}
F.~Rahaman, M.~Kalam, and A.~Ghosh, \emph{{Existence of wormholes in
  Einstein-Kalb-Ramond space time}}, Nuovo Cim. \textbf{121B} (2006), 303--307,
   \texttt{gr-qc/0605095}.

\bibitem{Mathur:2005zp}
S.~D. Mathur, \emph{The fuzzball proposal for black holes: An elementary
  review}, Fortsch. Phys. \textbf{53} (2005), 793--827,
  \texttt{hep-th/0502050}.

\bibitem{Greene:1995hu}
B.~R. Greene, D.~R. Morrison, and A.~Strominger, \emph{Black hole condensation
  and the unification of string vacua}, Nucl. Phys. \textbf{B451} (1995),
  109--120,  \texttt{hep-th/9504145}.

\bibitem{Hashimoto:2002sk}
K.~Hashimoto, P.-M. Ho, and J.~E. Wang, \emph{S-brane actions}, Phys. Rev.
  Lett. \textbf{90} (2003), 141601,  \texttt{hep-th/0211090}.

\bibitem{Hashimoto:2003qx}
K.~Hashimoto, P.-M. Ho, S.~Nagaoka, and J.~E. Wang, \emph{Time evolution via
  s-branes}, Phys. Rev. \textbf{D68} (2003), 026007,  \texttt{hep-th/0303172}.

\bibitem{Diaconescu:2000wy}
D.-E. Diaconescu, G.~W. Moore, and E.~Witten, \emph{{E(8) gauge theory, and a
  derivation of K-theory from M- theory}}, Adv. Theor. Math. Phys. \textbf{6}
  (2003), 1031--1134,  \texttt{hep-th/0005090}.

\bibitem{Loaiza-Brito:2006se}
O.~Loaiza-Brito, \emph{{Freed-Witten anomaly in general flux
  compactification}},  (2006),  \texttt{hep-th/0612088}.

\end{thebibliography}
\addcontentsline{toc}{section}{Bibliography}
\bibliographystyle{TitleAndArxiv}

\end{document}